# High-speed and high-resolution interrogation of FBG sensors using wavelength-to-time mapping and Gaussian filters

**Manuel P. Fernández,**[1,2,3,*] **Laureano A. Bulus Rossini,**[1,2,3] **, José L. Cruz,**[4] **Miguel V. Andrés,**[4] **and Pablo A. Costanzo Caso**[1,2,3,*]

[1] *LIAT, Comisión Nacional de Energía Atómica (CNEA), Bustillo 9500, Bariloche 8400 (RN), Argentina*
[2] *CONICET, CCT Patagonia Norte, Bariloche 8400 (RN) Argentina*
[3] *Instituto Balseiro (UNCuyo-CNEA), Bariloche 8400 (RN), Argentina*
[4] *Departamento de Física Aplicada y Electromagnetismo, Universidad de Valencia, Moliner 50, España*
*\*pcostanzo@ib.edu.ar; manuel.fernandez@ib.edu.ar*

**Abstract:** In this work we report a novel technique for simultaneous high-speed and high-resolution interrogation of fiber Bragg grating (FBG) sensors. The method uses the wavelength-to-time mapping effect in a chromatic dispersive medium and a couple of intensity Gaussian filters. The Bragg wavelength is retrieved by means of the amplitude comparison between the two filtered grating spectrums, which are mapped into a time-domain waveform. In this way, measurement distortions arising from residual power due to the grating sidelobes are completely avoided, and the wavelength measurement range is considerably extended with respect to the previously proposed schemes. We present the mathematical background for the interrogation of FBGs with an arbitrary bandwidth. In our proof-of-concept experiments, we achieved sensitivities of ~20 pm with ultra-fast rates up to 264 MHz.

## 1. Introduction

Bragg gratings are among the most popular optical devices and they find applications in a wide variety of fields such as optical signal processing [1,2], optical communications [3,4], microwave photonics [5], and sensoring [6,7]. In this sense, fiber Bragg grating (FBG) sensors are based on wavelength modulation, in which the sensed parameter (e.g. strain or temperature) is linearly related to the grating central wavelength [8].

The most extended FBG interrogation techniques use static filters to convert the wavelength shift into an intensity change. Among these methods, the edge filter [11-12], in which the detected light intensity is proportional to the wavelength drift, has attracted more attention due to its simple, low-cost and reliable structure. Interrogation of FBG sensors with speeds up to hundreds of megahertz is desirable in applications such as monitoring of ultra-fast dynamic phenomena, e.g. molecular dynamics sensing and in aerospace diagnostics. To this end, techniques based on the wavelength-to-time mapping using broadband short pulses and dispersive components have been demonstrated [13-16]. The main drawbacks of these schemes are the fundamental tradeoff between interrogation speed and wavelength resolution, and the need for expensive high-speed photodiodes and sampling electronics.

The conventional intensity-based methods require to be operated in the linear range of a filter to ensure a linear detection. Recently, Cheng et al [17] proposed an interrogation system in which the linear dependence is achieved using Gaussian filters. The difference between the intensities at the output of two crossed Gaussian filters ultimately leads to a linear behavior, which is exploited for the Bragg wavelength determination. This scheme has been proven to provide ultra-high sensitivity, and it is a promising solution for interrogation of FBGs with a wide spectrum [18]. However, the wavelength range in which the system shows a linear dependence is limited as power reflected from the grating sidelobes become non-negligible [18-19]. In fact, this shortcoming is common to every filtering-based techniques.

In this work, to solve this drawback, we propose and demonstrate a novel FBG interrogation technique based on crossed Gaussian filters, in which the filtered grating spectrum is mapped into a temporal waveform using a highly dispersive medium. The Bragg wavelength is then retrieved from the amplitude comparison between the time-mapped grating main lobes at the output of each filter. Consequently, measurement distortions arising from residual power due to the FBG side lobes are avoided, and the linear operational range is considerably extended. Moreover, in contrast to previous works in which the FBG spectral width must be either much narrower [17] or equal [18] to that of the Gaussian filters, we generalize the method for FBG sensors having an arbitrary spectral bandwidth, so the proposed system is further improved in terms of flexibility. Theoretical analysis and experimental results are presented and discussed.

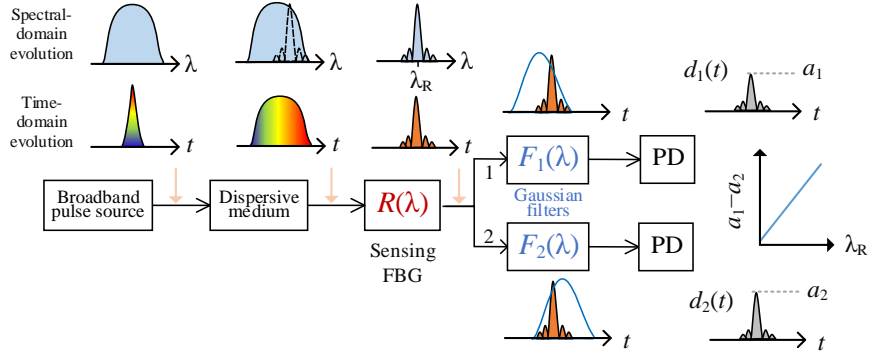

Fig. 1. Schematic representation of the FBG interrogation system based on the wavelength-to-time mapping and Gaussian filters.

## 2. Measurement principle

### 2.1. Mathematical background

Figure 1 illustrates the principle of the proposed FBG wavelength monitoring system. An optical source emits short broadband pulses at a repetition rate equal to the required interrogation speed. The pulses propagate through a highly dispersive medium with accumulated dispersion parameter $D$ (in ps/nm), such that the far-field Fraunhofer condition [20,21] is met at the output of the dispersive medium, and thus, the pulse spectrum is mapped into its temporal waveform with a wavelength-time relation given by the dispersion parameter. The resulting linearly-chirped pulse is used as probe pulse and transmitted to the sensing FBG, which is characterized by a reflectivity transfer function $R(\lambda)$. The reflected signal is then divided into two branches (branch 1 and 2) that present a couple of Gaussian filters with transfer functions $F_1(\lambda)$ and $F_2(\lambda)$. The signals at the output of the filters are detected and digitalized by separate photodiodes.

Assuming that the source spectrum is flat within the reflection spectrum of the FBG, and that the FBG has negligible chirp compared to that of the highly dispersive medium, the detected signals at the output of branches 1 and 2 can be expressed as

$$d_i(t) \propto R(\lambda')F_i(\lambda') \qquad i=1,2. \qquad (1)$$

where $\lambda' = t/D$ is the wavelength-time scale [20], [21]. Equation (1) implies that the overall system transfer function for each path, constituted by the product of the FBG and each Gaussian filter transfer functions, is mapped into the temporal waveform of the detected signals, with a time-wavelength scale determined by the accumulated dispersion $D$.

The Gaussian filters have both similar bandwidth $B_F$, and different central wavelength, $\lambda_{F1}$ and $\lambda_{F2}$. On the other hand, the main lobe of the FBG can be accurately described by a Gaussian function with bandwidth $B_R$ and central wavelength $\lambda_R$, especially for a low $\kappa L$ parameter [22], [23]. Thus, we can write the FBG and the Gaussian filters transfer functions, respectively, as

$$R(\lambda) \propto \exp\left(-4\ln(2)\left(\frac{\lambda - \lambda_R}{B_R}\right)^2\right) \quad \text{and} \quad F_i(\lambda) = \exp\left(-4\ln(2)\left(\frac{\lambda - \lambda_{Fi}}{B_F}\right)^2\right), \quad i = 1, 2. \quad (2)$$

By substituting Eqs. (2) into Eq. (1) we find that the detected waveform at the output of each branch $d_i(t)$ can be expressed as

$$d_i(t) = k \exp\left(-4\ln(2)\frac{(\lambda_R - \lambda_{Fi})^2}{B_R^2 + B_F^2}\right) \exp\left(-4\ln(2)\left(\frac{t - \tau_{di}}{T_d}\right)^2\right) \quad i = 1, 2, \quad (3)$$

where $k$ is a constant that includes the power of the probe pulse and different coefficients that are identical in both paths, e.g. propagation losses and splitting losses. The pulse full-width at half-maximum $T_d$ and its relative delay $\tau_{di}$ are given by

$$T_d = D\frac{B_R B_F}{\sqrt{B_R^2 + B_F^2}} \qquad \tau_{di} = D\frac{\lambda_R B_F^2 + \lambda_{Fi} B_R^2}{B_F^2 + B_R^2}. \quad (4)$$

Let us define the amplitude of the detected pulses of Eq. (3) in dB units as $a_i = 10\log(\max\{d_i(t)\})$, $i = 1, 2$. A linear dependence with the Bragg wavelength is achieved from the difference between the amplitudes of the detected signals in both channels. The amplitude comparison function (ACF) can be defined as the difference between both amplitudes, as follows

$$\text{ACF}(\lambda_R) = a_1 - a_2 = \frac{-40\log(e)\ln(2)\left(\lambda_{F1}^2 - \lambda_{F2}^2\right)}{B_R^2 + B_F^2} + \frac{80\log(e)\ln(2)\left(\lambda_{F1} - \lambda_{F2}\right)}{B_R^2 + B_F^2}\lambda_R. \quad (5)$$

From Eq. (5) it can be noted that the ACF presents a linear dependence with the Bragg wavelength $\lambda_R$. The slope of the ACF, which determines the systems' sensitivity, can be tuned by adjusting the wavelength offset of the Gaussian filters, $\lambda_{F1} - \lambda_{F2}$, and it also depends on the filters and grating bandwidths, $B_F$ and $B_R$.

Therefore, by means of a pulse-by-pulse digital signal processing, the grating Bragg wavelength can be retrieved from the measured pulses amplitude, $\tilde{a}_1$ and $\tilde{a}_2$, by applying the inverse amplitude comparison function $\lambda_R = \text{ACF}^{-1}(\tilde{a}_1 - \tilde{a}_2)$.

It is useful to compare the ACF of Eq. (5) with the difference between the Gaussian filters transfer functions (in dB). After some mathematical manipulations, it is straightforward to arrive to the following relation

$$\text{ACF}(\lambda_R) = \frac{B_F^2}{B_F^2 + B_R^2}\left(F_1(\lambda_R) - F_2(\lambda_R)\right). \quad (6)$$

Therefore, if the sensing FBG is much narrower than the Gaussian filters, such that $B_F^2 \gg B_R^2$, the amplitude comparison function is equal to the Gaussian filters difference. As the relative grating bandwidth increases, the slope of the ACF decreases. For instance, in the limit $B_F = B_R$, it is reduced to one-half of its maximum.

*2.2. Sidelobes consideration*

Power arising from the FBG residual sidelobes are the main factor limiting the linear operational range in the conventional measurement method, which involves the differential detection of the overall received power $P_i$, i.e. the integral of the product $R(\lambda)F_i(\lambda)$ [18,19]. Nevertheless, in the proposed method, the grating sidelobes are temporally separated from the main lobe in the detected signal due to the wavelength-to-time mapping, and thus, they do not contribute to the amplitude measurement. This allows to extend the linear measurement range, as exemplified in Fig. 2. For instance, Fig. 2(a) shows the Gaussian filters transfer functions where the wavelength offset is $\lambda_{F1} - \lambda_{F2} = 1.2$ nm and $B_F = 1.5$ nm, and a simulated spectrum of a uniform FBG with parameter $\kappa L = 0.6$, which yields a 3 dB bandwidth of $B_R = 415$ pm [22]. Figure 2(b) depicts the normalized spectrum at the output of the two branches when the FBG approaches the edge of one of the filters in a way that the residual sidelobes have non-negligible energy contribution to the output power. Figure 2(c) compares the amplitude difference $a_1 - a_2$ and the overall power difference $P_1 - P_2$ at the output of each branch as a function of the FBG central wavelength. It can be seen that the linear range is considerably extended using the amplitude comparison method, for which the linear trend is only broken when the residual sidelobes amplitude at the output of the Gaussian filters are higher than the main lobe amplitude. The slope of the function is found to be −11.93 dB/nm, in accordance to that expected from Eq. (5).

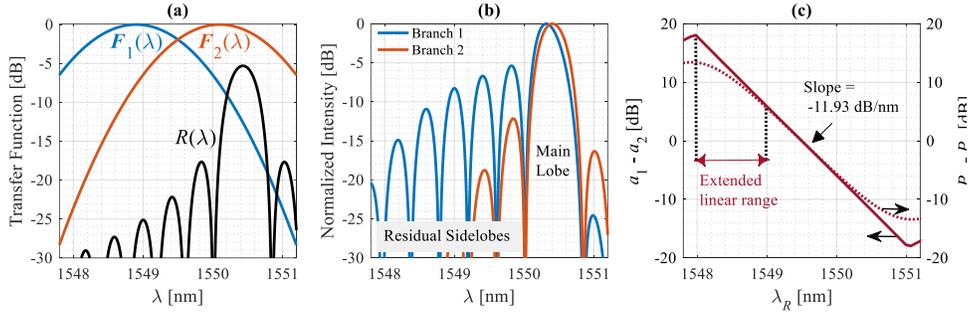

Fig. 2. (a) Transfer function of the Gaussian filters and the FBG reflectivity spectrum. (b) Spectrum of the signal at the output of each branch. (c) Comparison between the amplitude difference and the overall power difference.

## 3. Experiments and discussion

In order to probe the measurement principle and evaluate its performance, we implemented the experimental setup illustrated in Fig. 3. As optical source, we used a pulsed laser (Onefive Origami 15) that emits high peak-power picosecond pulses centered at 1550 nm at a tunable repetition rate. The pulses first propagate through a non-zero dispersion-shifter fiber spool in order to broaden their spectrum so it is approximately flat within the working wavelength range. Then, the pulses are attenuated and they propagate in the linear regime through the highly dispersive medium, which is a dispersion compensation module (DCM) with dispersion parameter $D = -1665.7$ ps/nm. The resulting linearly chirped pulses are amplified and launched to the sensing FBG through an optical circulator. A 3 dB coupler derives the reflected signal to a couple of tunable optical filters (TOF) with Gaussian transmittance and approximately equal bandwidths of $B_F = 1.4$ nm. The output signals are detected using similar photodiodes (Thorlabs DET08CFC) with a bandwidth of 5 GHz and digitized using a real-time oscilloscope (Rhode & Schwartz RTO 1044) with a bandwidth of 4 GHz.

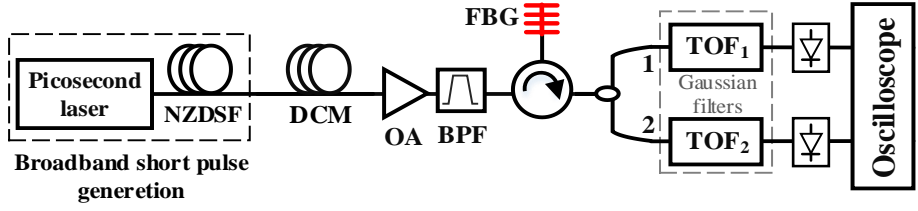

Fig. 3. Experimental setup of the FBG interrogation system. NZDSF: Non-zero Dispersion-shifted Fiber, DCM: Dispersion Compensation Module, OA: Optical Amplifier, BFP: Band-pass Filter, TOF: Tunable Optical Filter.

Firstly, we experimentally compared the achievable measurement range when a FBG is interrogated using the proposed method (based on the amplitude comparison $a_1 - a_2$) to that using the conventional method (based of the overall power difference $P_1 - P_2$). In our setup, the overall powers $P_i$ are obtained by substituting the photodetectors at the branches' output by two power meters. We obtained both the maximum amplitude difference and the power difference at the output of each branch as a function of the FBG central wavelength, whose results are shown in Fig. 4(a). In this case, the Gaussian filters are set to have wavelength offsets of 1.04 nm and 1.46 nm around a central wavelength of 1549.5 nm. The first thing to note in Fig. 4(a) is that the sensitivity, i.e. the slope of the function, is higher as the filter offset increases, as expected from Eq. (5). Moreover, it is seen that the linear trend is significantly extended when using the amplitude comparison method, since residual power reflected from the FBG sidelobes is avoided as it was expected from the numerical simulations (see Fig. 2(c)).

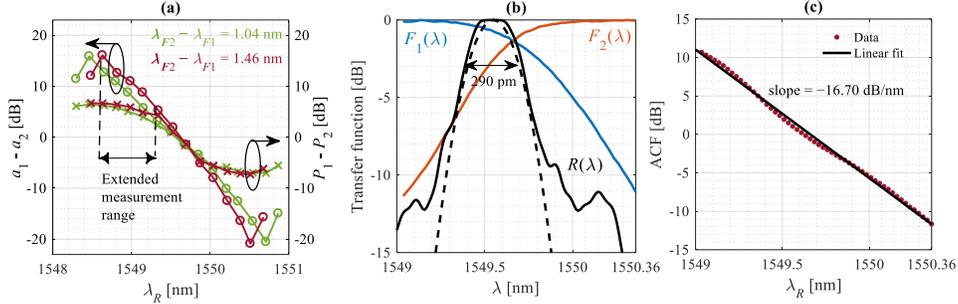

Fig. 4. (a) Linear wavelength range comparison for different wavelength offsets. (b) Measured Gaussian filters and FBG reflectivity spectrum. (c) Measured amplitude comparison function and linear fit.

In the following tests we fixed the filters' central wavelengths to $\lambda_{F1} = 1549$ nm (branch 1) and $\lambda_{F2} = 1550.3$ nm (branch 2). Therefore, the filters' wavelength offset is 1.3 nm. Figure 4(b) depicts the measured Gaussian filters transfer functions and the normalized reflectivity of the sensing FBG used in the experiments. The FBG main lobe is well adjusted by a Gaussian function with bandwidth $B_R = 0.29$ nm. The ACF for this setup is depicted in Fig. 4(c), which has been obtained from Eq. (6) by multiplying the measured filters difference data, $F_1 - F_2$, by the factor $B_F^2/(B_F^2 + B_R^2) = 0.96$. Together with the measured data it is plotted the corresponding linear fit, which presents a slope of −16.70 dB/nm.

The wavelength-to-time mapping of the filtered FBG spectrum at the output of the two Gaussian filters is shown in Fig. 5. Specifically, Figs. 5(a)-(b) depict the spectral and temporal waveforms at the output of branch 1, respectively, while Figs. 5(c)-(d) shows the equivalent signals at the output of branch 2. The wavelength-domain signals are measured by substituting the photodetector at the branches' output by an optical spectrum analyzer with a resolution of 40 pm. It can be appreciated that the temporal waveform presents a shape identical to that of its spectral counterpart, but temporally inverted due to the negative sign of the dispersion parameter. Superposed to the measured waveforms it is plotted the Gaussian fit of their main lobe. As expected, the temporal width of the fitting functions is equal in both channels, and it is found to be $T_d = 479$ ps. By substituting the pulse width into Eq. (4) the FBG bandwidth can be retrieved in order to use the appropriate ACF without the need for a previous grating characterization stage. Moreover, although the residual sidelobes have non-negligible energy contribution (which is especially appreciated at branch 2 in Fig. 5), in the presented method the FBG sidelobes are temporally separated from the main lobe, and thus they do not contribute to the measurement, i.e. the pulse amplitude.

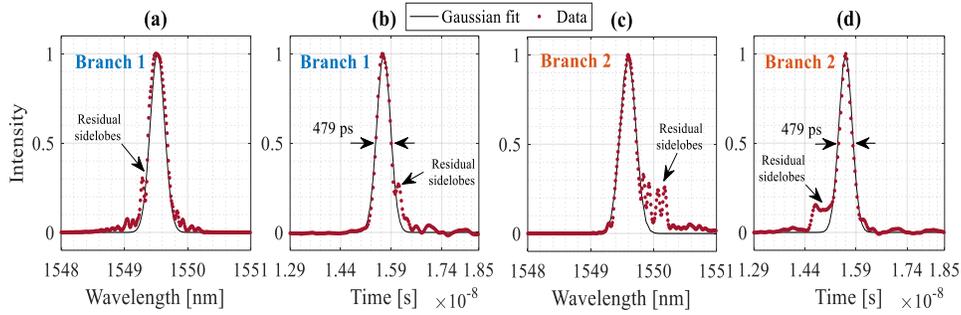

Fig. 5. Normalized waveforms at the output of (a)-(b) branch 1 and (c)-(d) branch 2.

The system performance was tested through the estimation of the central wavelength of the FBG sensor when it was swept from an initial value of 1549.4 nm to a final value of 1549.7 nm. The amplitude comparison function shown in Fig. 4(c) was employed to map the measured pulse amplitude difference into the estimated Bragg wavelength. The measurement results are shown in the scatter plot of Fig. 6, where it is represented the measured Bragg wavelength versus the actual one, together with the residual measurement error. It is seen that the obtained measurement error is significantly low and the absolute error does not exceed 20 pm. This wavelength measurement error also depends on the uncertainty of the physical parameters in the case of a temperature/strain sensing system. For instance, the grating sensitivity coefficients are measured to be 8 pm/°C and 1.2 pm/µε. Thus, a maximum deviation of 20 pm in the wavelength estimation leads to a maximum error for temperature and strain sensing of 2.5 °C and 16.6 µε, respectively.

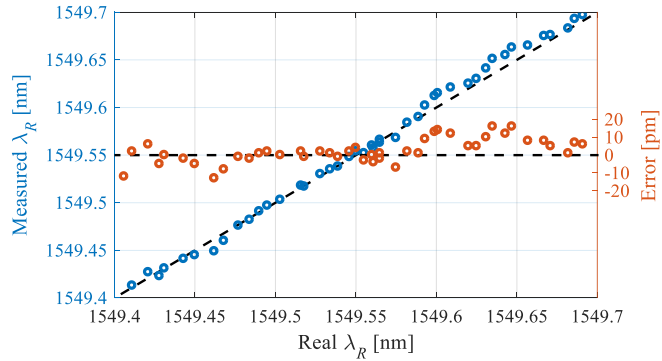

Fig. 6. Wavelength estimation results and residual error.

The interrogation speed is determined by the laser repetition rate *R*, which should be kept such that there is no temporal overlap between two consecutive pulses even in the pessimistic case in which the FBG experience the maximum but opposite wavelength shifts. In simultaneous to an amplitude variation in the detected waveforms, the Bragg wavelength shift leads to a temporal shift proportional to the system's dispersion. This can be appreciated in Fig. 7, where the detected signals at the output of both branches under different Bragg wavelength shifts ranging from an initial value $\lambda_0$ to $\lambda_0 + 294$ pm, are shown. In the experiments, the interrogation period was set to 3.78 ns, yielding an interrogation rate of 264 MHz. The temporal shift experienced by the filtered pulses is $\tau_{di}$, given by Eq. (4). Thus, the repetition period of the probe laser should be higher than the maximum temporal shift difference, $\max(\tau_{di}) - \min(\tau_{di})$, to which it must be added a guard interval that considers pulse width, residual side lobes and detectors relaxation times. In this case, potential interrogation speeds up to the GHz regime can be reached, while maintaining high resolutions of a few picometers.

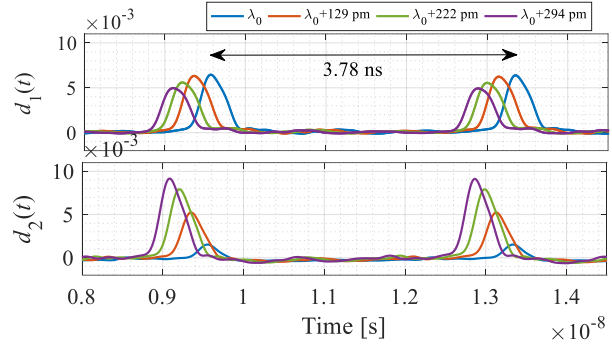

Fig. 7. Acquired signal for different Bragg wavelength shifts. The laser repetition rate is set to 264 MHz.

## 4. Conclusions

We have proposed a novel FBG wavelength monitoring technique based on Gaussian filters and wavelength-to-time mapping of broadband short pulses in a dispersive medium. In the proposed system, the filtered grating spectrum is mapped into a temporal waveform, and the Bragg wavelength is determined from the amplitude difference of the detected pulses at the output of two crossed Gaussian filters. Therefore, in contrast to previous schemes, we increase the linear operational range by avoiding measurement distortions arising from residual power from the sensor side lobes. Also we have demonstrated the mathematical background for a general case in which the sensing FBG can have an arbitrary spectral width. Moreover, we proved that the grating bandwidth can be directly derived from the detected waveform signal, which further improves the system in terms of flexibility and real-time operation. Finally, we carried out proof-of-concept experiments and obtained measurement errors of less than 20 pm in simultaneous with ultra-fast acquisition rates of 264 MHz.

### Acknowledgments

This paper was partially supported by CNEA, ANPCyT, CONICET, UNCUYO and UV. M.P.F. is a fellow of CONICET; P.A.C.C. and L.A.B.R. are professors at IB and researchers of CONICET; and J.L.C and M.V.A. are professors and researchers of UV.